# Dielectrophoretic Assembly of High-Density Arrays of Individual Graphene Devices for Rapid Screening

Aravind Vijayaraghavan,[1,*] Calogero Sciascia,[2,3] Simone Dehm,[1] Antonio Lombardo,[3] Alessandro Bonetti,[3] Andrea C. Ferrari,[3,†] and Ralph Krupke[1, 4, §]

[1] Institut für Nanotechnologie, Forschungszentrum Karlsruhe, 76021 Karlsruhe, Germany; [2] INFM-CNR Physics Department, Politecnico di Milano, Milano, Italy; [3] Engineering Department, University of Cambridge, Cambridge CB3 0FA, United Kingdom; [4] DFG Center for Functional Nanostructures (CFN), 76028 Karlsruhe, Germany

Address correspondence to: * v.aravind@int.fzk.de, † acf26@eng.cam.ac.uk, § krupke@int.fzk.de

TOC GRAPHIC

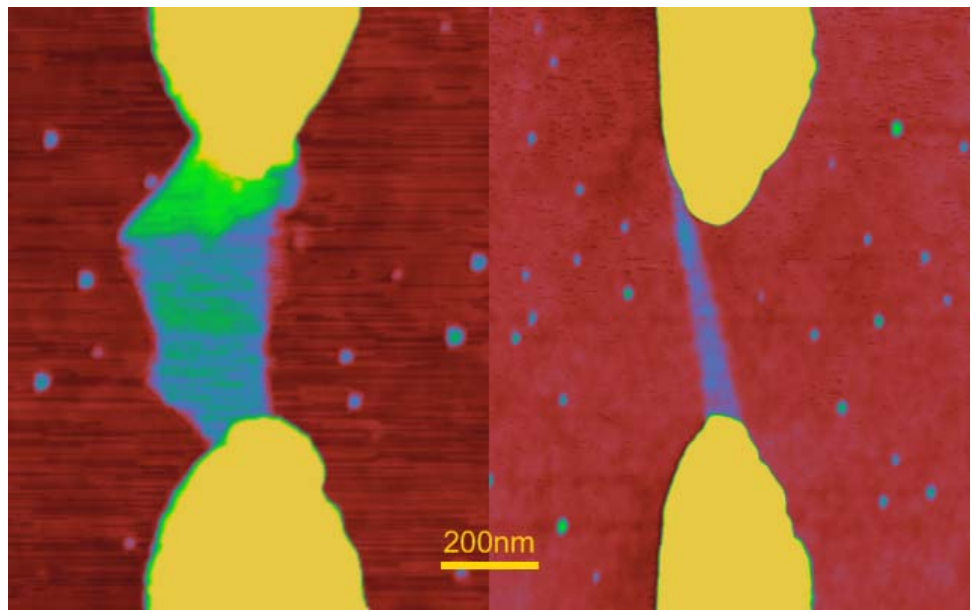

ABSTRACT

We establish the use of dielectrophoresis for the directed parallel assembly of individual flakes and nanoribbons of few-layer graphene into electronic devices. This is a bottom-up approach where source and drain electrodes are prefabricated and the flakes are deposited from a solution using an alternating electric field applied between the electrodes. These devices are characterized by scanning electron microscopy, atomic force microscopy, Raman spectroscopy and electron transport measurements. They are shown to be electrically active and their current carrying capacity and subsequent failure mechanism is revealed. Akin to carbon nanotubes, we show that the dielectrophoretic deposition is self-limiting to one flake per device and is scalable to ultra-large-scale integration densities, thereby enabling the rapid screening of a large number of devices.

INTRODUCTION

Graphene, as a free-standing, two-dimensional crystal of carbon atoms, was experimentally shown to exist only in 2004[1], although for the past 60 years it has been theoretically studied as the fundamental structural and electronic building block for various $sp^2$ bonded carbon allotropes such as graphite,[2] fullerenes and carbon nanotubes (CNTs).[3,4] Graphene's promising applications in nanoelectronics[5] have triggered a rush of research into its production and integration into functional electronic components. Multi-layered graphene reaches the 3D limit of graphite in terms of its electronic and dielectric properties at about 10 layers[6,7] and therefore the term 'few-layer graphene (FLG)' will be used in this paper within this limit. Electronic devices based on graphene can be fabricated either in a top-down approach, where graphene is first grown or deposited on a substrate and subsequently contacted by electrodes, or a bottom-up approach where the desired electrodes are prefabricated on a substrate and graphene subsequently self-assembles or is incorporated at the device locations.

In the top-down approach, the most popular method to generate graphene for research purposes is the micromechanical cleavage of bulk graphite.[1] However, this is a low-yield process and monolayer graphene flakes have to be discerned from a majority of thicker flakes, and is therefore unsuitable for controlled or scaled-up device fabrication. Graphene can also be grown by chemical vapor deposition from hydrocarbon sources on metal substrates[8-16] or by thermal annealing of SiC.[17, 18] Metal substrates are unsuitable for electronic device applications and require sample transfer to insulating substrates in order to make useful devices,[13, 16] while the SiC route inherently limits the substrate choice. An alternative route is to start from graphene solutions. Graphene has been randomly deposited from suspension[19] on to substrates, the flakes located by AFM or SEM and electrodes fabricated on top to contact them to yield functional graphene devices. Such a procedure is not easily scalable for controlled device fabrication. Directed assembly of graphene flakes at predetermined locations is thus required.

Here, we demonstrate the fabrication and characterization of electronically active devices of individual FLG flakes using the bottom-up approach, where the flakes are selectively deposited from solution only in between pre-defined electrodes in a high-density array using dielectrophoresis. One approach to obtain graphene solutions involves the dispersion of graphene oxide (GO),[20-22] stabilized by hydroxyl and epoxide functionalization. GO can be subsequently reduced to graphene,[23] but leaves a significant number of defects that disrupt the electronic properties.[24] Recently, much progress has been made in the chemical exfoliation of graphene from bulk graphite. Graphene dispersions, with concentrations of up to 0.01 mg/ml, have been produced by exfoliating graphite in organic solvents such as N-methyl-2-pyrrolidone (NMP).[25] Graphene dispersions so obtained could be further processed by techniques such as density gradient ultracentrifugation to isolate single-layer and multilayer graphene with high separation yield. Graphene nanoribbons[19, 26] have been fabricated by acid treatment of intercalated expandable graphite and subsequent dispersion and sonication.

Dielectrophoresis has emerged as a powerful technique for the controlled fabrication of nanoelectronic devices.[27, 28] Recently, thin-film devices of GO soot particles[29] and epitaxial-graphene – GO junctions[30] have been fabricated by dielectrophoresis. Dielectrophoresis has been applied to separate metallic and semiconducting CNTs[31] and for the fabrication of thin CNT films with controlled alignment and properties.[32, 33] We have recently demonstrated that individual CNTs can be assembled at ultra-large integration densities into functional electronic devices using dielectrophoresis.[34] Here, we successfully adopt a similar approach for the fabrication of scalable arrays of functional, individual FLG devices in a three-terminal configuration, although the dimensionality of the FLG flakes is different from that of nanotube or nanowires. This method holds various advantages over other routes for graphene device fabrication and allows for rapid screening of a large number of flakes and devices.

RESULTS AND DISCUSSIONS

Fig. 1 shows a representative region of the device array, where 11 out of 15 devices are successfully bridged by a flake. A zoom in to 5 devices is shown in Supporting Information. The thickness of the flakes and number of layers is subsequently confirmed by AFM and Raman spectroscopy measurements. We observe flakes of graphite (Figs. 2(a, b) and FLG (Figs. 2(d, e)) and FLG nanoribbons (Figs. 2(g, h)). SEM images of other such devices can be found in Supporting Information. We observe that in SEM, suspended graphite and FLG sections show brighter contrast compared to the substrate, while flakes laying flat on the substrate show similar contrast to the substrate irrespective of the number of layers and are identifiable primarily based on edge-contrast. Even within a single flake, regions of different thicknesses can only be discerned in the SEM if the edge-contrast is substantial. SEM imaging was performed at 10keV acceleration voltage, in order to minimize surface-charging that might perturb the electronic properties of graphene, as has been reported in the case of CNTs.[35, 36] In the absence of charging induced contrast mechanisms, the secondary electron emission coefficients of $SiO_2$ and C (graphene/graphite) are nearly identical.[37] In the case of CNTs, it has been proposed that charging of a suspended CNT in interaction with the electron beam causes large local electric fields around it which results in an enhanced secondary electron emission.[38] Similar contrast enhancement or suppression can be obtained by applying an appropriate external bias to the CNT.[39] In an alternate mechanism, it was experimentally shown that a fast electron beam passing through a CNT can excite surface plasmons.[40] These can excite and accelerate electrons which give an enhanced secondary electron emission probability to the CNT,[38] and a similar effect might also exist for suspended graphene. Considering the high electron energies, high conductivity of the FLG and the low contact resistance (as shown later), we propose that the latter mechanism is the likely cause for enhanced contrast of suspended flakes.

A high trapping efficiency was obtained with parameters previously optimized for single-wall CNT device assembly.[34] Further optimization of field frequency and amplitude and graphene concentration is expected to improve the device yield. Note that elongated flakes and nanoribbons assemble with their long axis along the connecting line of the electrodes, as is expected for the induced dipole moment. The quality of the deposited FLG flakes is directly related to the contents of the source suspension. We expect that arrays deposited from a suspension consisting predominantly of single-layer or bi-layer graphene or graphene nano-ribbons will give significantly higher yield of devices of the same.

As described earlier,[34] the density of electrodes on the surface is limited by the thickness of the insulating oxide and integration densities of 1 million devices per $cm^2$ are obtained. Dielectrophoretic deposition is seen to be self-limiting to one flake or nanoribbon in each device location, because of the higher polarizability of the deposited FLG compared to the surrounding medium. This is similar to previous results with CNTs, and a similar mechanism can also be expected here. When the first such flake or nanoribbon is deposited in the electrode-gap, it changes the dielectrophoretic force fields in its vicinity from attractive to repulsive, thereby limiting further deposition in that electrode gap. In thick graphite devices, which have low resistance (shown later), the short-circuiting of the floating electrode with the grounded electrode might also contribute to the self-limiting assembly.

More than 50 flakes were analyzed by Raman spectroscopy. A majority of them consist of multilayer graphene, with some double-layer flakes. We did not find graphene monolayers for the present solutions. On one hand, we expect less than 1% of the flakes to be monolayers under the sonication conditions used here.[25] On the other hand, dielectrophoretic force on the flakes scales proportionally to the volume, and thereby, thicker flakes are deposited preferentially by this process. Figure 3 shows the Raman spectrum obtained with a 514 nm excitation, of four deposited flakes of increasing number of layers, from bi-layer to thick graphite, as evidenced by the shape of the Raman 2D peak.[41, 42] In particular, the four sub-bands are a clear indication of bilayer graphene.[42] A D peak is present, which we attribute to the flake edges due to the smaller size of the flakes compared to the excitation laser spot.[41-43]

Finally, we show that the devices fabricated here are electronically functional. Thick graphite and FLG flakes show linear *IV* characteristics with low resistances of <10 kΩ (Fig. 2(c)). Thin FLG flakes have slight non-linearity at low bias (Fig. 2(f)), while the thin nanoribbon (bi-layer) shows a pronounced low-bias current suppression (Fig. 2(i)). The scaling of non-linearity with number of layers was previously reported on FLG obtained by reduction of GO.[44] This was attributed to conductance suppression in the first graphene layer, owing to interactions with the substrate. The layers would behave as parallel conductors with negligible inter-layer conduction, implying that contribution of the first layer diminishes as the flake thickness increases. Raman spectroscopy, however, does not reveal any substrate doping effects in our devices, i.e. no up-shift and broadening of the G peak.[41, 45-47] It has also been proposed that when a graphene monolayer is deposited on an oxygen terminated $SiO_2$ surface, it exhibits a band-gap opening.[48] This gap reduces as the number of deposited layers increases (as in FLG). Since our $SiO_2$ substrate was subject to oxygen plasma treatment prior to deposition, as was required to enable wetting of the NMP, the observed non-linearity could be due to this graphene-substrate interaction. Further investigations, such as deposition of graphene on hydrogen terminated surfaces and the use of other solvents that do not require oxygen plasma treatment of the surface, are currently underway. However, we were unable to detect gate-bias dependence of transconductance in any of the samples. This can be attributed to two factors. The gate-bias window to observe expected ambipolar behavior in graphene has been seen to be as high as ±50 V for 300 nm thick gate-oxide.[1] The 800 nm thick gate-oxide in our devices makes this range even wider due to weaker gate coupling. Also, the charge neutrality point is often shifted beyond the ±20 V gate range used in our measurements. Due to the presence of unbridged floating electrodes on the surface, which capacitively couple to the back-gate potential, we are unable to explore a gate voltage range wider than ±20 V without an electric-discharge breakdown between the floating and grounded electrodes.

We note that the resistance of the graphene devices increases upon electron irradiation and decreases after a high-current annealing procedure,[49] similar to CNT devices.[35, 36] An increase in resistance of up to 3 orders of magnitude was reported in CNT devices due to the perturbing effect of the high local electric fields arising from charges implanted in the substrate in the vicinity of the nanotube. When these charges are drained through the nanotube under a high bias, the resistance recovers to its original value. However, the changes in current due to electron irradiation and subsequent recovery are substantially smaller (less than an order of magnitude) in graphene compared to CNTs. Detailed investigation of these phenomena in the case of graphene will be presented elsewhere. Our FLG devices were able to sustain high currents, greater than 10 µA/layer. This represents current densities of $10^7$ A/cm$^2$. High-current failure is seen to occur always at the graphene-metal contact, often involving local melting of the metal electrodes, as shown in Fig. 4 and Supporting Information. This suggests that the failure mechanism in our devices involves joule-heating of the graphene-metal contact or some other thermally-assisted failure mechanism. The electrode melting might also be attributed to electromigration in the narrow Au-graphene contact region at such high current densities. The region of the FLG flake adjacent to the contact is suspended in most cases and not effectively thermalized by the substrate. It is therefore expected to be the hottest region. It is known in the case of CNTs the nanotube temperature can exceed the melting point of Au (1064 ºC) at high currents[50] and a similar mechanism might be in effect in our graphene devices leading to the melting-failure of the electrode. We do not reach the regime of current saturation before failure, where the generation of non-thermalized 'hot' phonons would be the dominant failure mechanism, as in CNTs.[51] The critical current density is also an order of magnitude less than for completely substrate supported graphene,[1, 52] owing to the suspended portion of the FLG flake adjacent to our electrodes.

CONCLUSIONS

In summary, we have shown that dielectrophoretic deposition enables rapid assembling of individual graphene devices into high-density arrays with high yield. It holds a number of advantages over other methods of graphene device fabrication. Since NMP is used as the solvent the FLG flakes are not coated with any surfactant layer. It is scalable to ultra-large scale integration densities and is self-limiting to one flake or nanoribbon per device. It avoids high-temperature processing steps and is compatible with existing microelectronic fabrication technologies. The method is independent of the graphene source and an improvement of the suspensions, in yield and layer selectivity via density gradient ultracentrifugation or similar techniques, could allow the fabrication of high-density arrays of single-layer or bi-layer graphene or graphene nanoribbons. The graphene flakes can also be subsequently patterned to form nano-ribbons or other branched-ribbon architectures. Such patterning does not require any previous AFM or SEM imaging of the flakes, since their location and the orientation of the electrodes is predefined. We expect that dielectrophoretic deposition of graphene from suspension will emerge as widely used method for device fabrication for both research and commercial purposes.

MATERIALS AND METHODS

FLG flakes are dispersed in organic solutions following a similar procedure to that described in Ref. 22. The starting graphite flakes (NGS Naturgraphit GmbH) have an area of 0.1 mm$^2$ to few mm$^2$. NMP (Sigma-Aldrich) is used as the organic solvent, as it has been found effective in the case of CNTs in forming solutions without a surfactant.[53-55] 5 mg of graphite was dissolved in 10 ml of NMP, sonicating for 30 minutes followed by centrifugation for 30 minutes at 1000 rpm and 20 ºC.

FLG devices are fabricated on a substrate of degenerately doped Si with an 800 nm insulating SiO$_2$ surface layer. The electrodes are first defined by electron-beam lithography and consist of 40 nm Au over a 5 nm Ti adhesion layer. An electrode array design, similar to that for CNTs,[34] is adopted for FLG devices (Fig. 1). It consists of one common drain electrode, which is biased, and an array of floating independent source electrodes, which are not directly connected to the A/C source. The alternating electric field is applied between the common drain and conducting Si back-gate. Both the source and the drain electrodes can also be directly biased, however this limits the scalability of the process. Instead, all the floating source electrodes capacitively couple to the gate and acquire a similar potential to it. Prior to deposition, the substrate is rendered hydrophilic by an oxygen plasma treatment to enable the NMP to wet it. A drop (20 µl) of suspension is then placed on the substrate and an alternating electric field of 300 kHz and 2 V/μm is applied for 3 mins. The suspension is subsequently removed by a $N_2$ flow.

The devices are imaged by a LEO1530 Scanning Electron Microscopy (SEM) and Digital Instruments Multimode Atomic Force Microscopy (AFM) in tapping-mode to characterize the deposited flakes. The deposited flakes were also characterized by Raman spectroscopy and imaging using a Renishaw and a Witec spectrometer, respectively. Electron transport measurements were performed with nanoprobes mounted on Kleindiek Nanotechnik MM3A-EM micromanipulators in-situ in the SEM.

ACKNOWLEDGMENTS

The authors acknowledge C. W. Marquardt, M. Engel, J. Coleman, Y. Hernandez and M. Lotya for helpful discussions. A.V., S.D. and R.K. acknowledge funding by the Initiative and Networking Fund of the Helmholtz-Gemeinschaft Deutscher Forschungszentren (HGF); A.CF. acknowledges funding from the Royal Society and European Research Council grant NANOPOTS, C. S. from "Fondazione CARIPLO" and Politecnico di Milano, A. L. from Palermo University.

SUPPORTING INFORMATION AVAILABLE

SEM images of the high-density array and individual graphene devices at higher magnification and SEM and AFM images of graphene device failure. This material is available free of charge via the Internet at http://pubs.acs.org.

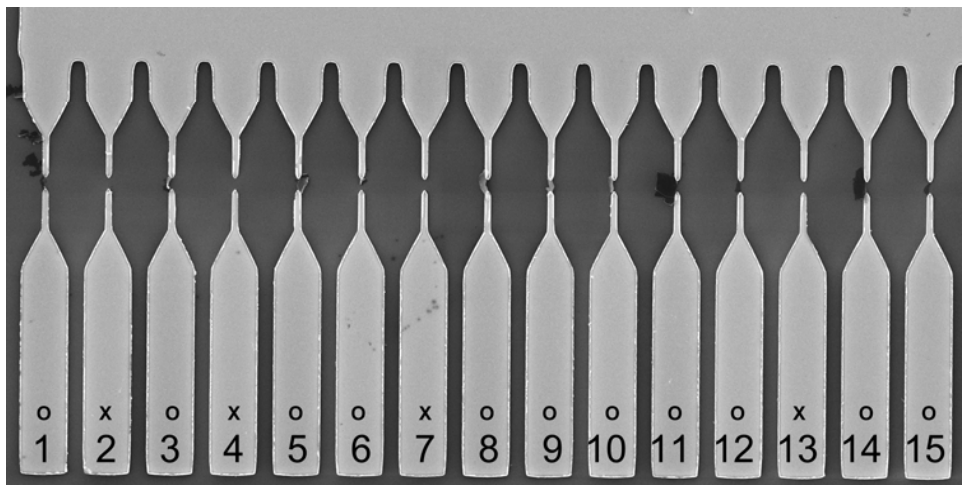

**Figure 1.** Scanning electron micrograph of a representative region of an array of Graphene devices fabricated by dielectrophoretic deposition from a suspension in NMP. Each device comprises of a floating source electrode (bottom, labeled 1 – 15), a common drain electrode (top) and a common back-gate. 11 out of the 15 devices in this region contain a graphene flake located between the electrode tips. Successfully bridged electrodes are marked as **o** while nonfunctional devices are marked as **x**. A zoom in to 5 of these devices is presented as supporting information.

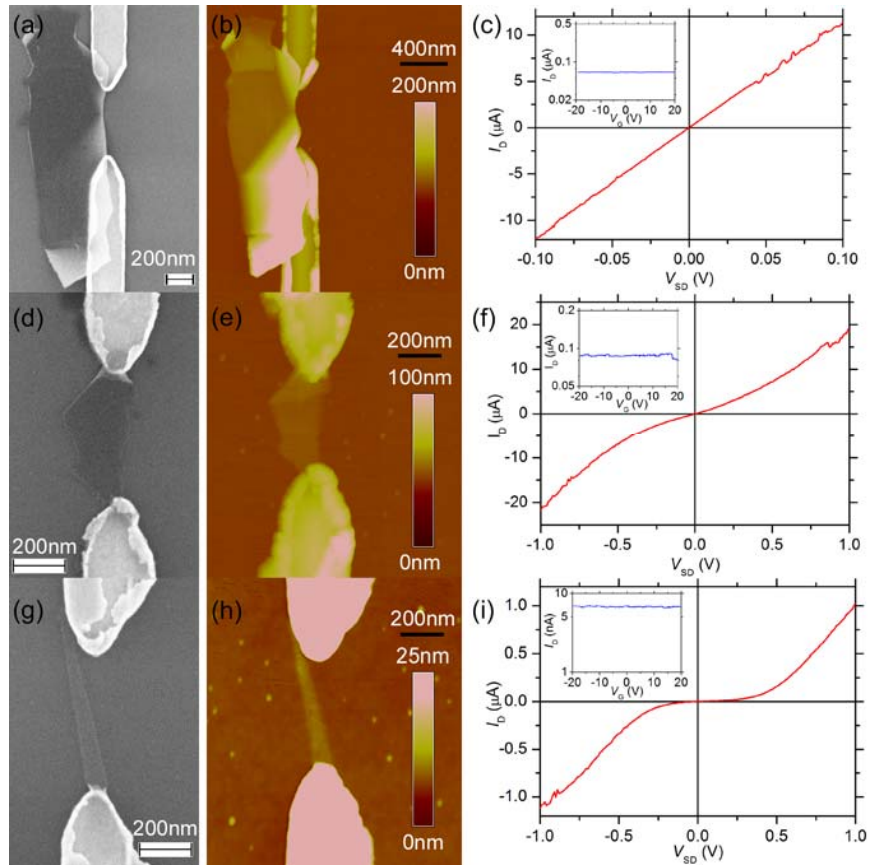

**Figure 2. (a, d, g)** Scanning electron micrograph, **(b, e, h)** atomic force micrograph and **(c, f, i)** transport measurements on different devices. **(a – c)** Graphite flake, showing linear *IV* characteristics. **(d – f)** Thin FLG flake (~5 layers, ~3nm thick), showing slight low-bias current suppression. **(g – i)** Thin graphene nanoribbon (~2 layers, ~1.5nm thick, ~60nm wide) showing pronounced low-bias current suppression.

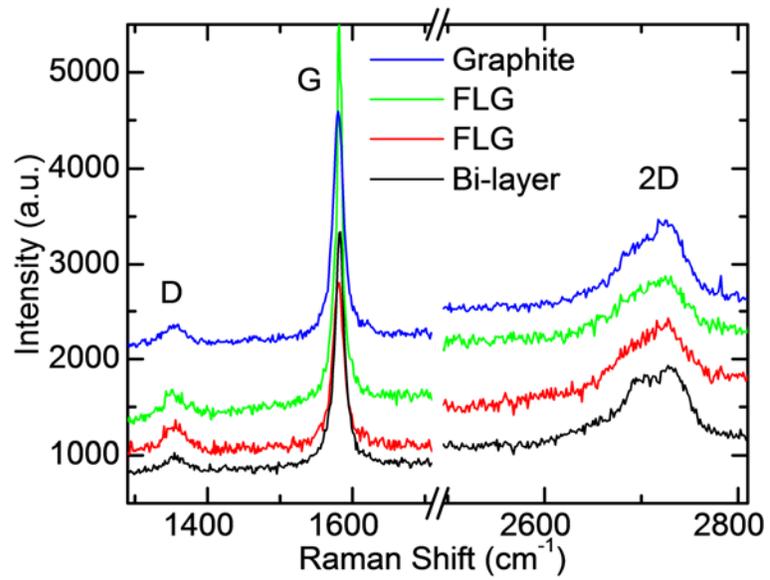

**Figure 3.** Raman spectra of 4 flakes. The number of layers is determined by the shape of the 2D peak, in combination with AFM height measurements.

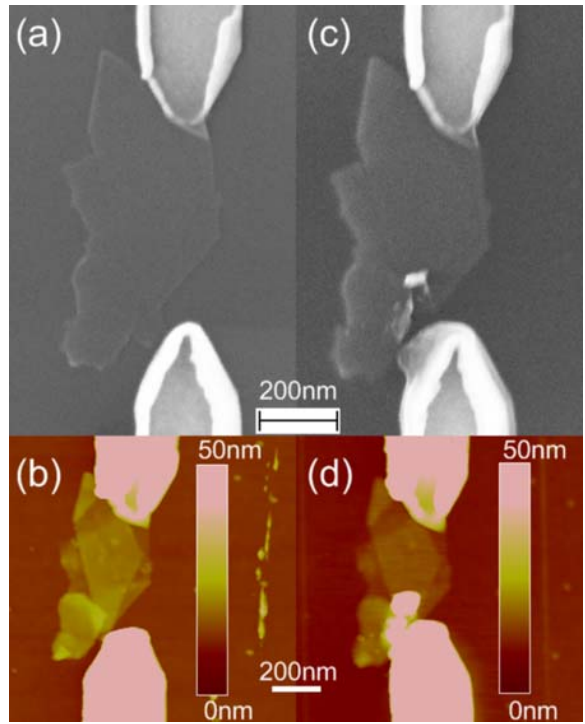

**Figure 4. (a, c)** SEM and **(b, d)** AFM of a graphene device **(a, b)** before and **(c, d)** after high-current failure. Melting of the metal at the graphene – electrode contact is evident. Supporting information contains additional such images.

# Dielectrophoretic Assembly of High-Density Arrays of Individual Graphene Devices for Rapid Screening – Supporting Information

Aravind Vijayaraghavan,[1,*] Calogero Sciascia,[2,3] Simone Dehm,[1] Antonio Lombardo,[3] Alessandro Bonetti,[3] Andrea C. Ferrari,[3,†] and Ralph Krupke[1, 4, §]

[1] Institut für Nanotechnologie, Forschungszentrum Karlsruhe, 76021 Karlsruhe, Germany; [2] INFM-CNR Physics Department, Politecnico di Milano, Milano, Italy; [3] Engineering Department, University of Cambridge, Cambridge CB3 0FA, United Kingdom; [4] DFG Center for Functional Nanostructures (CFN), 76028 Karlsruhe, Germany

Address correspondence to: * v.aravind@int.fzk.de, † acf26@eng.cam.ac.uk, § krupke@int.fzk.de

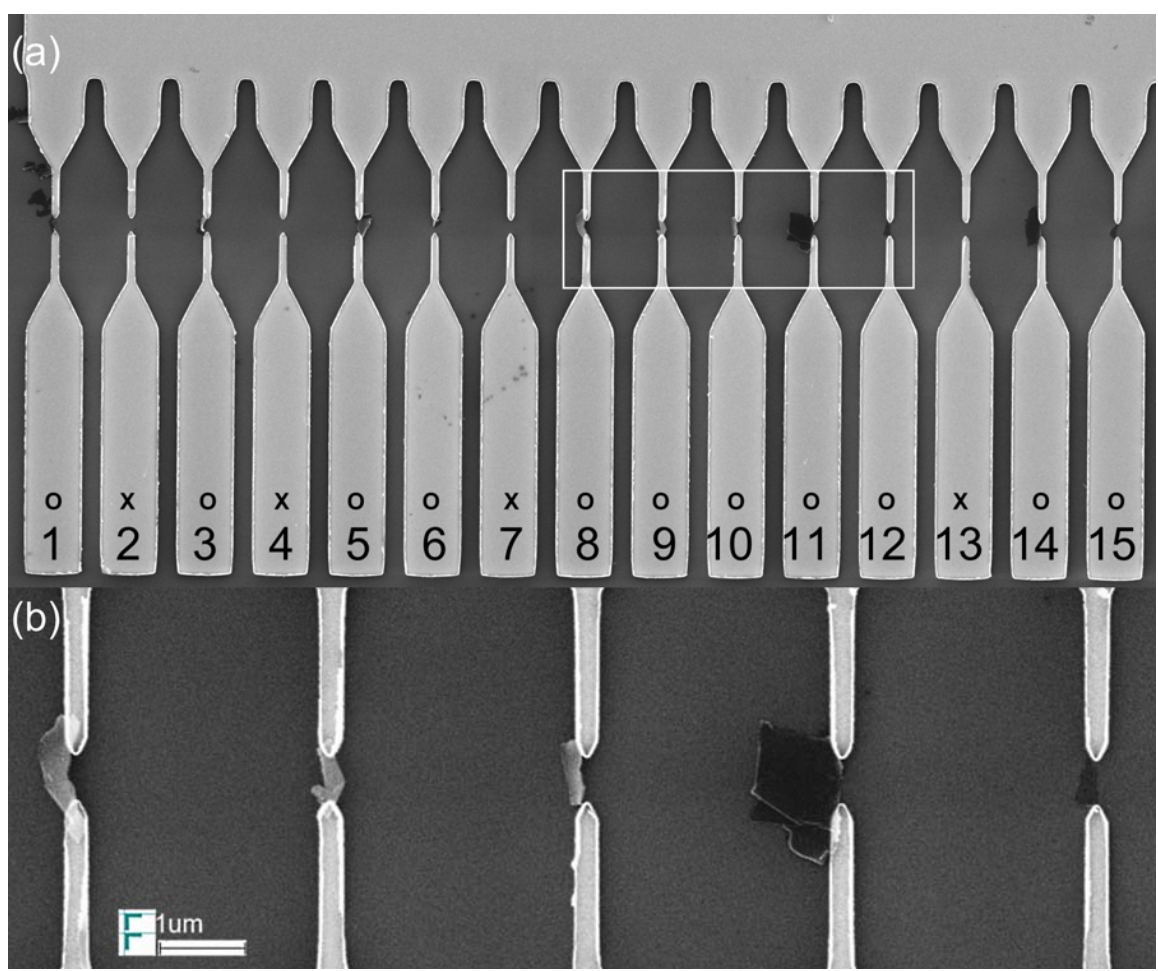

**Supporting Figure 1. (a)** Scanning electron micrograph of a representative region of an array of Graphene devices fabricated by dielectrophoretic deposition from a suspension in NMP. Each device comprises of a floating source electrode (bottom, labeled 1 – 15), a common drain electrode (top) and a common back-gate. 11 out of the 15 devices in this region contain a graphene flake located between the electrode tips. Successfully bridged electrodes are marked as o while nonfunctional devices are marked as x. **(b)** Zoom in to 5 of these devices showing the graphene flakes bridging the electrode gaps.

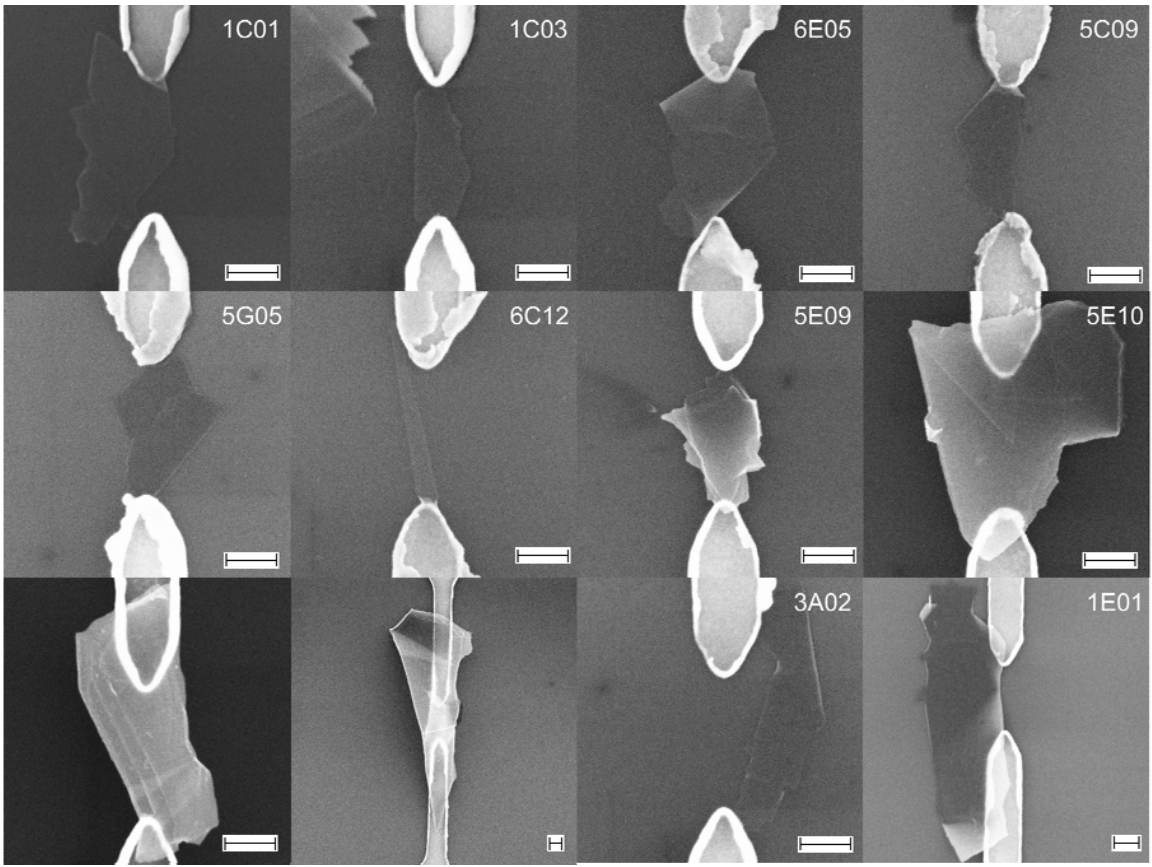

**Supporting Figure 2.** SEM images of 12 representative FLG devices of various thicknesses (layers) and sizes, including a nanoribbon (6C12). This distribution of shapes and sizes of the flakes reflects their distribution in the source suspension, and a suspension consisting of predominantly one kind of graphene, such as single-layers, bi-layers or nanoribbons will yield devices of the same.

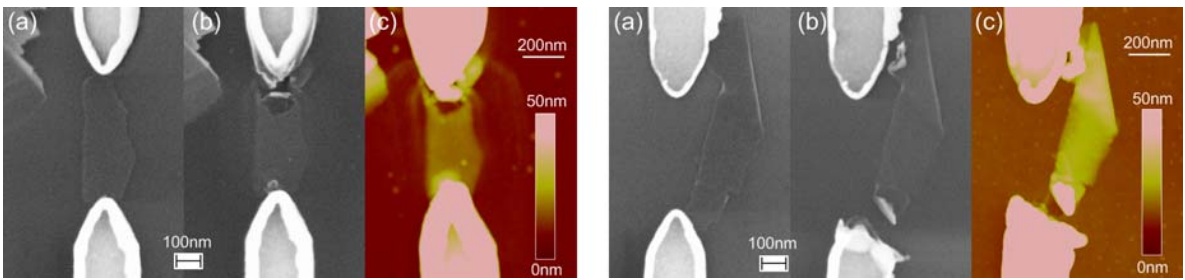

**Supporting Figure 3.** Two FLG devices **(a)** before and **(b, c)** after high-current failure. Failure is seen to occur at the FLG-electrode contact involving local melting of the metal.